\documentclass[a4paper,12pt]{article}

\usepackage[affil-it]{authblk}
\usepackage{amsmath}
\usepackage{amssymb}
\usepackage{graphicx}
\usepackage{subfig}
\usepackage{bbm}
\usepackage{epsfig}
\usepackage{yfonts}

\def\({\left(} \def\){\right)}
\def\[{\left[} \def\]{\right]}

\def\hz{\mathrel{\mathop h^{\scriptscriptstyle{(0)}}}{}\!\!}

\def\gz{\mathrel{\mathop g^{\scriptscriptstyle{(0)}}}{}\!}
\def\go{\mathrel{\mathop g^{\scriptscriptstyle{(1)}}}{}\!}
\def\hz{\mathrel{\mathop h^{\scriptscriptstyle{(0)}}}{}\!}
\def\ho{\mathrel{\mathop h^{\scriptscriptstyle{(1)}}}{}\!}

\def\gt{\,\tilde{g}\,}
\def\gs{\mathrel{\mathop g^{\scriptscriptstyle{(2)}}}{}\!\!\!}

\def\hz{\mathrel{\mathop h^{\scriptscriptstyle{(0)}}}{}\!\!}

\def\gz{\mathrel{\mathop g^{\scriptscriptstyle{(0)}}}{}\!}
\def\go{\mathrel{\mathop g^{\scriptscriptstyle{(1)}}}{}\!}

\newcommand{\be}{\begin{equation}}
\newcommand{\ee}{\end{equation}}
\newcommand{\bea}{\begin{eqnarray}}
\newcommand{\eea}{\end{eqnarray}}
\newcommand{\ba}{\begin{eqnarray}}
\newcommand{\ea}{\end{eqnarray}}

\newcommand{\beq}{\begin{equation}}
\newcommand{\eeq}{\end{equation}}
\newcommand{\beqa}{\begin{eqnarray}}
\newcommand{\eeqa}{\end{eqnarray}}
\newcommand{\beqar}{\begin{eqnarray*}}
\newcommand{\eeqar}{\end{eqnarray*}}

\begin{document}

\title{Spacetime Entanglement with $f(\mathcal{R})$ Gravity}
\author{Razieh Pourhasan\footnote{e-mail: {\tt
rpourhasan@perimeterinstitute.ca}}}

\affil{Perimeter Institute for Theoretical Physics, 31 Caroline
St. N., Waterloo, ON, N2L 2Y5, Canada}

\affil{Department of Physics \& Astronomy, University of Waterloo,
Waterloo, Ontario N2L 3G1, Canada}

\maketitle

\begin{abstract}
We study the entanglement entropy of a general region in a theory
of induced gravity using holographic calculations. In particular
we use holographic entanglement entropy prescription of
Ryu-Takayanagi in the context of the Randall-Sundrum 2 model while
considering general $f(\mathcal{R})$ gravity in the bulk. Showing
the leading term is given by the usual Bekenstein-Hawking formula,
we confirm the conjecture by Bianchi and Myers for this theory.
Moreover, we calculate the first subleading term to entanglement
entropy and show they agree with the Wald entropy up to extrinsic
curvature terms.
\end{abstract}
\newpage

\section{Introduction}
\label{intro}

Dividing a quantum system into two parts, {\it entanglement
entropy} (EE) is a measure of quantum correlations between the two
subsystems. EE has appeared in different areas such as condensed
matter theory \cite{Gang:2006}, quantum information theory
\cite{oai:2225, oai:4163} and black hole physics
\cite{Bombelli:1986, oai:3048, oai:2145}. In QFT, the EE between a
spatial region $A$ and its complement, at fixed time, is given by
the usual von-Neumann entropy: $S=-tr[\rho_A\log\rho_A]$. Whereas
the reduced density matrix $\rho_A$ is obtained by tracing out the
degrees of freedom in the complementary region and describes the
remaining degrees of freedom in region $A$. Note that, because of
the appearance of divergences in continuum limit, one needs to
introduce a UV regulator to make sense of this calculation.

\begin{figure}[t!]
\center{\vspace{-1cm}
\includegraphics[width=300pt]{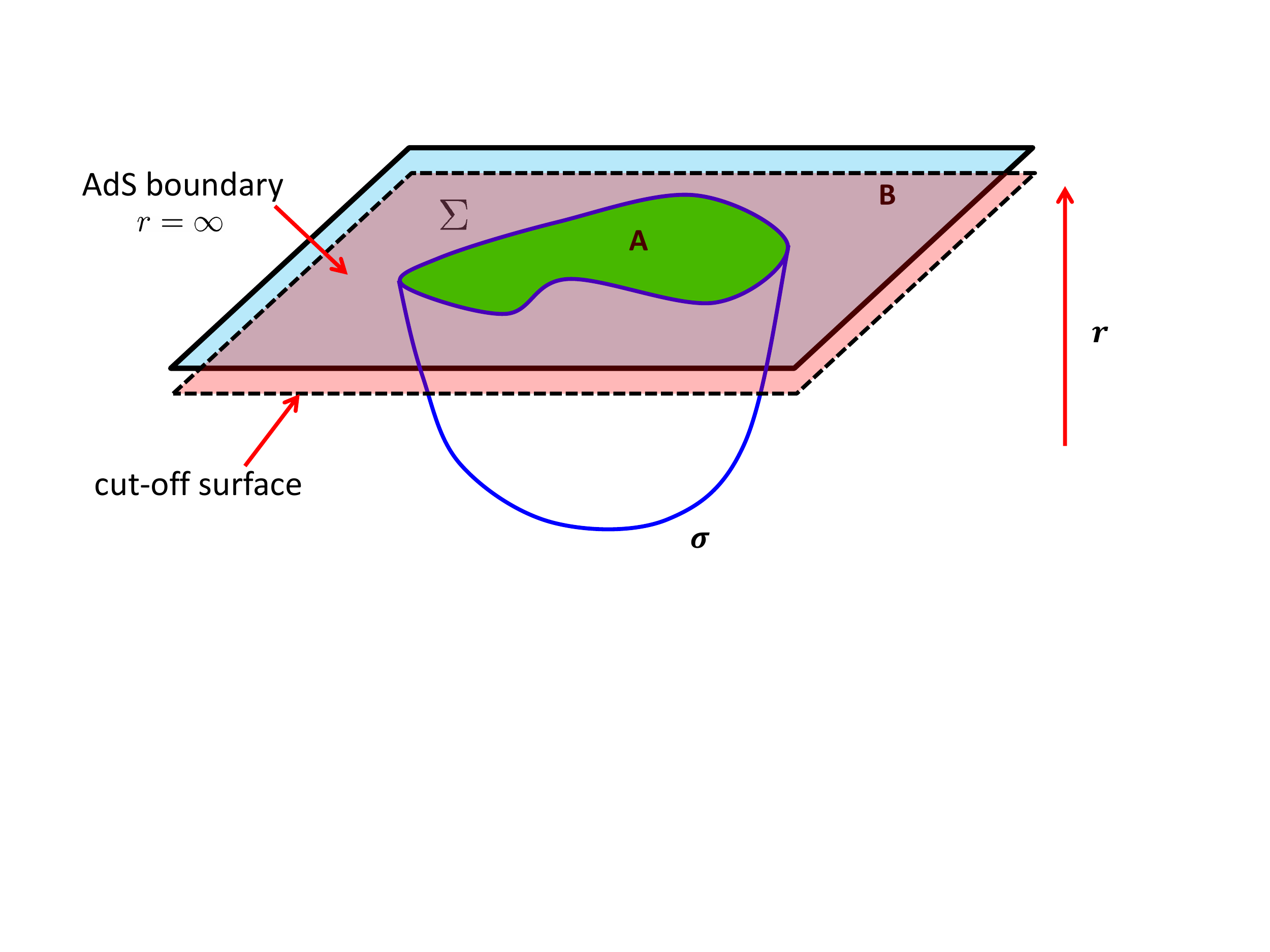}\vspace{-2cm}}
    \caption{Ryu-Takayanagi surface $\sigma$ associated to the entangling surface $\Sigma$ on the AdS boundary.}\label{fig:minsurf}
\end{figure}

More recently, EE has been widely studied in the AdS/CFT, after an
elegant conjecture by Ryu and Takayanagi \cite{Ryu:2006bv,
Ryu:2006ef} which relates EE in the boundary to the classical
geometry in the bulk through holography. According to this
prescription, the EE associated with an entangling surface
$\Sigma$ in the CFT, i.e., on the conformal boundary of an AdS
bulk, is determined by evaluating $A/4G$ on an extremal surface
$\sigma$ in the bulk, which extends to the AdS boundary to meet
$\Sigma$. Here the UV boundary cut-off corresponds to introducing
a regulator surface at some finite large $r$, as shown in figure
(\ref{fig:minsurf}). While the conjecture has passed lots of
consistency tests \cite{Ryu:2006bv, Hung:2011xb, Casini:2011kv,
Headrick:2010zt}, e.g., reproduces the same results for
2-dimensional CFT and for the thermal ensemble as well as
connection to central charges of CFT, it has been proved recently
by Lewkowycz and Maldacena \cite{Lewkowycz:2013nqa}.

In a related framework, Randall and Sundrum showed that the
standard four-dimensional gravity will arise at long distances on
a brane embedded in a warped five-dimensional background
\cite{Randall:1999ee, Randall:1999vf}. One may construct such a
model in arbitrary dimension by taking two copies of
$(d+1)$-dimensional AdS spacetime, gluing them together along a
cut-off surface at some large radius while inserting $(d-1)$-brane
at this junction.

Here we are motivated with the recent conjecture by Bianchi and
Myers proposing that in quantum gravity, EE of a general surface
with smooth geometry in a smooth background is finite and the
leading term is the usual Bekenstein-Hawking formula. Therefore,
in this letter, which is a followup for the previous work with
Myers and Smolkin \cite{Myers:2013lva}, we first briefly introduce
the Randall-Sundrum 2 (RS2) model in section \ref{warped}. In
section \ref{fRg}, we use the RS2 model to derive induced gravity
on a $(d-1)$-brane which is embedded in the $(d+1)$-dimensional
bulk with $f(\mathcal{R})$ gravity theory. Then in section
\ref{EntEnt}, we use holographic entanglement entropy to study EE
of a general surface on this brane. Finally, we summarize the
results in section \ref{disc}.

\section{Warped Spacetime}
\label{warped}

If string theory is expected to be the UV completion of general
relativity, then the existence of extra dimensions is required in
order to cure quantum anomalies from stringy gauge symmetries
\cite{Ooguri:1996ik}. Although the size and shape of the extra
dimensions can not be predicted by string theory, they need to be
compactified and sufficiently small so that 4-dimensional
spacetime is recovered at low energies. One way to do this is
through \textit{warped compactification}, as proposed by Randall
and Sundrum \cite{Randall:1999vf}. Here, we focus on a version
known as the RS2 model.\footnote{There is also the RS1 model
\cite{Randall:1999ee} in which there exists another brane in the
interior of the bulk, i.e., the IR brane, located at a finite
distance from the UV brane. Therefore, one can think of the RS2 as
a limit of the RS1 where the IR brane has been taken to infinity.}
In this set-up a single 3-brane is embedded in a five dimensional
bulk given by
\begin{equation}
I_{bulk}=\frac{1}{16\pi G_5}\int d^4x\int dr\sqrt{-g}
\left(\frac{12}{L^2}+\mathcal{R}\right)\,,\label{RSaction}
\end{equation}
where $L$ is the scale of cosmological constant as well as the
curvature radius of $AdS_5$ spacetime and $r$ is the fifth extra
dimension in a bulk. Therefore, one can produce a 4D effective
theory on 3-brane from 5D gravitational action in the bulk.
According to UV/IR duality, this brane is usually called the UV
brane, since in the bulk we are working in the IR regime of
gravity theory by considering the region at smaller radius. In
fact, in addition to the bulk action there exists the 4D action of
the brane which to leading order is the brane tension, looking
like the cosmological term contribution, and any other matter
fields living on the brane. Indeed, one can think of the 4D metric
as it is multiplied by a \textit{warp factor} which is an
exponential function of the fifth coordinate, i.e., locally we
have the AdS geometry as
\begin{equation}
ds^2=e^{-2r/L}\eta_{\mu\nu}dx^{\mu}dx^{\nu}+dr^2\label{RSmetric}
\end{equation}
where $L$ is assumed to be large compared to the 5D Planck scale,
indicating that the bulk is smoothly curved. Therefore we can
trust that metric (\ref{RSmetric}), which is the solution to the
5D Einstein equation of the total action, i.e., bulk plus brane,
is a valid solution provided the tension of the brane is set to be
$T_{brane}=3/4\pi L G_5$.
\begin{figure}[t!]
\begin{center}
\includegraphics[width=.7\textwidth]{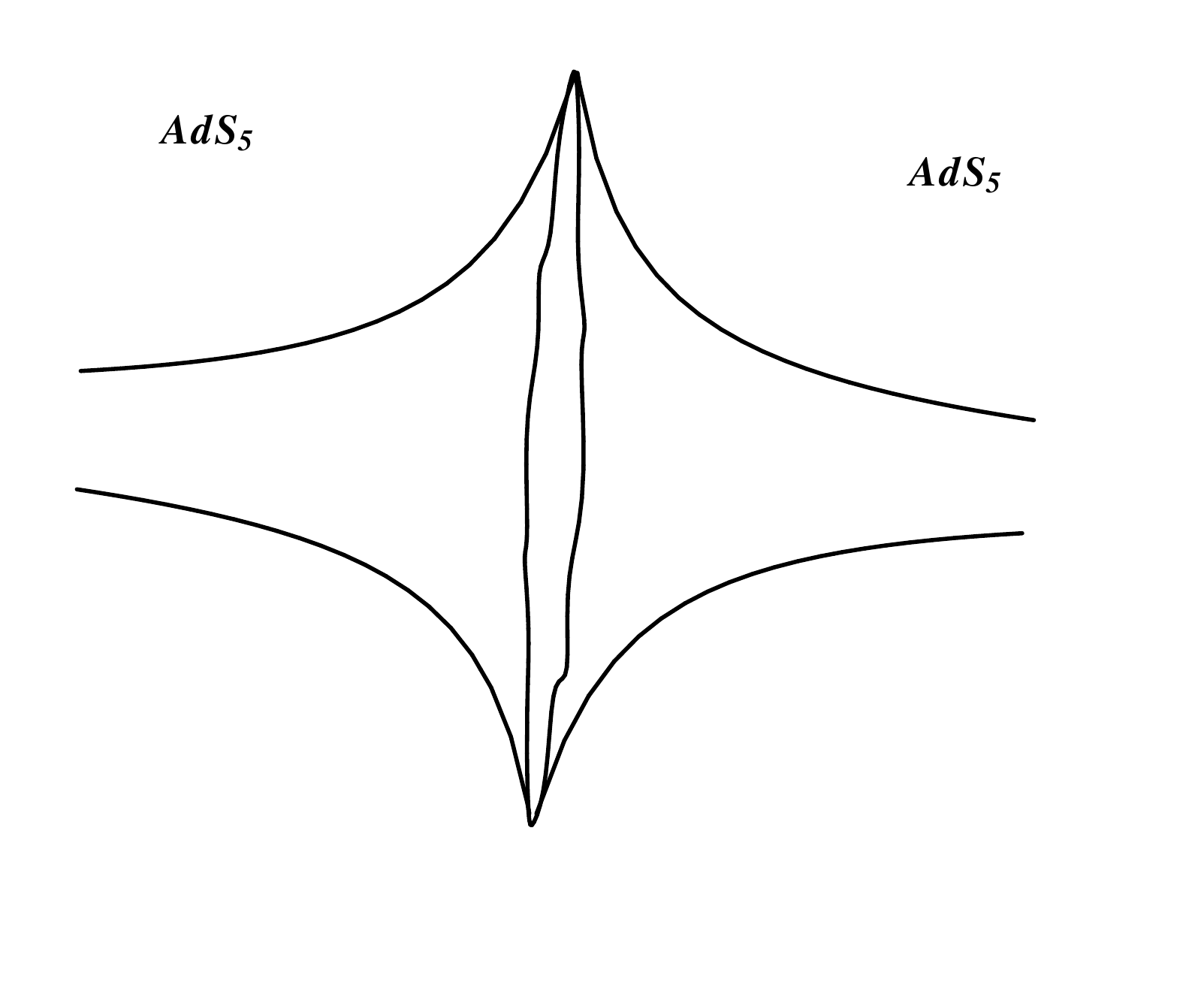}
\end{center}
\caption{Two copies of $AdS_5$ spaces have been glued together at
some finite radial direction and a 3-brane has been inserted at
the junction.}\label{AdS}
\end{figure}

The most important outcome of the model is the 4D effective
Newton's constant in terms of the bulk parameters as
\begin{equation}
G_N=\frac{2G_5}{L}
\end{equation}
which is obtained from 5D action by integrating out the extra
dimension. Thus in the RS2 model, gravity is localized in the
vicinity of the brane which contains 4D UV theory with finite
effective coupling. One can easily extend this construction to
arbitrary dimensions to describe a $d$-dimensional UV theory on
the brane from a $d+1$-dimensional AdS bulk.

In the rest of this letter, we use the RS2 construction as a
holographic framework to study the entanglement entropy of a
general region in induced gravity, i.e., the 4D gravity theory on
the brane. To do so, we assume two copies of AdS spacetime, cut
each of them at some finite radial distance, glue them together at
the cut and then insert a brane at this junction-- see figure
(\ref{AdS}). Therefore according to the RS2 mechanism, we have
localized gravity in the vicinity of the brane and induced gravity
on the brane with finite coupling. On the other hand, due to
AdS/CFT correspondence, we have two copies of strongly coupled CFT
living on the brane coupled to induced metric of the brane. Also
cutting the AdS geometry at some finite radius corresponds to
introducing a finite UV regulator in the dual field theory. Since
this cutting surface is also the location of the brane, in order
to deal with one length scale on the brane, without loss of
generality, we choose the AdS scale in the bulk to be equal to the
UV cut-off on the boundary theory where we denote both with
$\delta$. The fact that the QFT cut-off matches the AdS curvature
scale indicates the bulk geometry is highly curved if $\delta$ is
small.

\section{$f(\mathcal{R})$ gravity in the bulk}
\label{fRg}

We will consider extended theories of gravity in the bulk as a
generalization of the usual Einstein-Hilbert action
(\ref{RSaction}). In particular, we study the RS2 model with
$f(\mathcal{R})$ gravity where $f$ is an arbitrary function of the
Ricci scalar \cite{Sotiriou:2008rp}, as an interesting toy-model.
Thus the action for the $AdS_{d+1}$ bulk becomes
\begin{equation}
I_{bulk}=\frac{1}{16\pi G_{d+1}}\int d^dx d\rho
\sqrt{-G}\bigg[\frac{d(d-1)}{L^2}+f(\mathcal{R}(G))\bigg]+I_{surf}\label{bulkaction}
\end{equation}
where $G_{d+1}$ is the gravitational constant in the bulk, $L$ is
the scale of cosmological constant and $\mathcal{R}$ is the
curvature scalar in the bulk. The dimensionless coordinate $\rho$
is the extra radial direction in the bulk and $x^{\mu}$ are the
coordinates along the brane located at $\rho=\rho_c$ whereas
$\rho=0$ would be the boundary of $AdS_{d+1}$.\footnote{The
dimensionless coordinate $\rho$ is related to dimensionfull
coordinate $r$ in previous section as $\rho=e^{-2r/\delta}$.} Note
that the $AdS_{d+1}$ geometry again has the curvature radius
$\delta$ matching the cut-off in the boundary theory. However, we
will see that $AdS_{d+1}$ scale is no longer the cosmological
scale, i.e., $\delta\neq L$. To have a well-defined action the
proper surface term is of the form \cite{Dyer:2008hb}
\begin{equation}
I_{surf}=\frac{1}{8\pi G_{d+1}}\int
d^dx\sqrt{-\tilde{g}}\,\mathcal{K}\,
f'(\mathcal{R})|_{\rho=\rho_c}\,,\label{surf}
\end{equation}
where $\mathcal{K}$ is the trace of second fundamental form of the metric on the
brane and prime denote derivative with respect to $\mathcal{R}$.

We use Fefferman-Graham gauge \cite{Fefferman:2007rka} for the
metric in the bulk which is
\begin{equation}
ds^2=G_{\mu\nu}dx^{\mu}dx^{\nu}=\frac{\delta^2}{4}\frac{d\rho^2}{\rho^2}+\frac{1}{\rho}g_{ij}(x,\rho)dx^idx^j\,,\label{metric}
\end{equation}
where $\delta$, the curvature radius of $AdS_{d+1}$, is related to
the cosmological constant $L$ and the gravitational couplings
implicit in $f(\mathcal{R})$ through the equation of motion in the
bulk, i.e.,
\begin{equation}
f'(\mathcal{R})\mathcal{R}_{\mu\nu}+\bigg(G_{\mu\nu}\nabla^{\sigma}\nabla_{\sigma}-\nabla_{\mu}\nabla_{\nu}\bigg)
f'(\mathcal{R})-\frac{G_{\mu\nu}}{2}\bigg(f(\mathcal{R})+\frac{d(d-1)}{L^2}\bigg)=0\,.\label{eom}
\end{equation}
That is, if one inserts the metric (\ref{metric}) with
$g_{ij}=\eta_{ij}$, i.e., pure AdS space, into (\ref{eom}) one
obtains
\begin{equation}
\frac{1}{L^2}=-\frac{1}{d(d-1)\delta^2}\bigg[2df'(\mathcal{R}_0)+\delta^2f(\mathcal{R}_0)\bigg]\,,
\label{Lvsdelta}
\end{equation}
where $\mathcal{R}_0$ is the curvature of $AdS_{d+1}$ spacetime,
i.e.,
\begin{equation}
\mathcal{R}_0=-\frac{d(d+1)}{\delta^2}\,.\label{R0}
\end{equation}

One can obtain the induced gravity action on the brane by
integrating out the extra radial dimension of the bulk action
(\ref{bulkaction}). To do so, we use a derivative expansion for
the metric $g_{ij}$ about the position of the brane of the form
\begin{equation}
g_{ij}(x,\rho)=\overset{\scriptscriptstyle{(0)}}{g}_{ij}+\rho
\overset{\scriptscriptstyle{(1)}}{g}_{ij}+\rho^2
\overset{\scriptscriptstyle{(2)}}{g}_{ij}+\cdots\,,\label{derexp}
\end{equation}
where $\overset{\scriptscriptstyle{(0)}}{g}_{ij}$ is the metric of
the $AdS$ boundary at $\rho=0$ and
\begin{eqnarray}
\overset{\scriptscriptstyle{(1)}}{g}_{ij} &=&
-\frac{{\delta}^2}{d-2}\bigg(R_{i j}[\gz\,]-
\frac{\overset{\scriptscriptstyle{(0)}}{g}_{ij}}{2(d-1)}R[\gz\,]
\bigg)\,,\\
\gs_{i j} &=& \delta^4\,\bigg(k_1\,  C_{m n k l}C^{m n k l}\gz_{i
j} +
k_2\, C_{i k l m}C_{j}^{~~k l m} \nonumber \\
&& + \frac{1}{d-4}\bigg[\frac{1}{8(d-1)}\nabla_i\nabla_jR -
\frac{1}{4(d-2)} \Box R_{i j}+
\frac{1}{8(d-1)(d-2)} \Box R \gz_{i j} \nonumber\\
&&-\frac{1}{2(d-2)}R^{k l}R_{i k j l} +
\frac{d-4}{2(d-2)^2}R_{i}^{~k}R_{j k}+
\frac{1}{(d-1)(d-2)^2}RR_{i j}\nonumber \\
&&+\frac{1}{4(d-2)^2}R^{k l}R_{k l}\gz_{i j}
-\frac{3d}{16(d-1)^2(d-2)^2}R^2 \gz_{i j}\bigg]\bigg)\,,
\label{metricexpand2}
\end{eqnarray}
with $R_{ij}$ and $C_{mnkl}$ being the Ricci and Weyl tensors
associated with the boundary metric
$\overset{\scriptscriptstyle{(0)}}{g}_{ij}$, respectively
\cite{Imbimbo:1999bj}. The two constants $k_1$ and $k_2$ depend on
the type of gravity theory in the bulk. By solving the equation of
motion (\ref{eom}) for $f(\mathcal{R})$ in the bulk, one
explicitly finds $k_1,\,k_2=0$. The latter are most easily
determined if one picks a fixed geometry on the boundary for
$\overset{\scriptscriptstyle{(0)}}{g}_{ij}$ and then plugs the
metric expansion (\ref{derexp}) into (\ref{eom}).

Using the metric expansion (\ref{derexp}) one can perform a
derivative expansion for the curvature scalar in the bulk; it is a
matter of calculation to find\footnote{The details of calculation
could be found in \cite{Myers:2013lva}.}
\begin{equation}
\mathcal{R}=\mathcal{R}_0+\cdots\,,\label{rexpansion}
\end{equation}
since we are just interested in the terms up to curvature squared,
we don't really need to specify ellipsis which are of
$\mathcal{O}(\partial^6)$ and higher. Indeed, as it is manifestly
shown in \cite{Myers:2013lva}, the only curvature squared term in
the expansion (\ref{rexpansion}) has a coefficient depending on
the constants $k_1$ and $k_2$. However, this term is absent in the
present case with $f(\mathcal{R})$ gravity for which $k_1$ and
$k_2$ are both zero.

According to scaling symmetry of AdS the position of the brane
could be set to $\rho=\rho_c=1$ without loss of generality,
therefore we need to justify that the derivative expansion
(\ref{derexp}) can be reasonably truncated for finite $\rho$. It
could be realized from (\ref{metricexpand2}) that truncation can
be consistently achieved by demanding that the boundary metric
$\overset{\scriptscriptstyle{(0)}}{g}_{ij}$ is weakly curved on
the scale of $AdS$ curvature $\delta$. More precisely, we require
\begin{equation}
\delta^2 R^{ij}_{~~kl}[\gz]\ll1.\label{deltaR}
\end{equation}
and similarly for covariant derivatives of the curvatures. Note
that to make sense of this inequality, we are assuming that the
expressions are evaluated in an orthonormal frame. Although
$\overset{\scriptscriptstyle{(0)}}{g}_{ij}$ is not the metric on
the brane, the constraint (\ref{deltaR}) is sufficient for the
brane to be also smoothly curved. It is evident from the
expression for the induced metric on the brane
\begin{equation}
\gt_{ij}=G_{ij}|_{\rho=1}=\overset{\scriptscriptstyle{(0)}}{g}_{ij}+
\overset{\scriptscriptstyle{(1)}}{g}_{ij}+
\overset{\scriptscriptstyle{(2)}}{g}_{ij}+\cdots\label{gbrane}\,,
\end{equation}
that the difference between two metrics is small, i.e.,
$\gt_{ij}-\overset{\scriptscriptstyle{(0)}}{g}_{ij}\simeq\overset{\scriptscriptstyle{(1)}}{g}_{ij}\ll1$
when (\ref{deltaR}) holds.

In order to calculate the induced gravity action on the brane
which is given by
\begin{equation}
I_{ind}=2I_{bulk}+I_{brane}\,,\label{indact}
\end{equation}
where the factor of two for the bulk gravitational action is due
to the fact that we have two copies of AdS spaces and $I_{brane}$
accounts for any contribution from matter fields localized on the
brane as well as brane tension. However, here for simplicity we
focus only on the latter, i.e.,
\begin{equation}
I_{brane}=-T_{brane}\int d^d x \sqrt{-\gt}\,.\label{braneaction}
\end{equation}

We also need to find the derivative expansions for the extrinsic
curvature $\mathcal{K}_{ij}$ at the brane where the
outward-pointing unit normal vector is given by
$n_{\mu}=-\sqrt{G_{\rho\rho}}\delta^{\rho}_{\mu}$. Therefore, one
can easily derive
\begin{eqnarray}
\mathcal{K}_{ij}=\nabla_{i}n_{j}|_{\rho=1}=-\frac{\rho}{\delta}\frac{\partial
G_{ij}}{\partial\rho}|_{\rho=1}=\frac{1}{\delta}\bigg(\gt_{ij}-\sum_{n=1}^{\infty}n\overset{\scriptscriptstyle{(n)}}{g}_{ij}\bigg)\,,
\end{eqnarray}
which up to curvature squared terms yields to
\begin{equation}
\mathcal{K}=\frac{1}{\delta}\bigg[d+\frac{\delta^2}{2(d-1)}R
+\frac{\delta^4
}{2(d-1)(d-2)^2}\bigg(R_{ij}R^{ij}-\frac{d}{4(d-1)}R^2\bigg)\bigg]+\mathcal{O}(\delta^6)\,.
\end{equation}
Note that curvatures in the above expression are constructed from
the brane metric $\gt_{ij}$.

Finally putting together (\ref{indact}), (\ref{bulkaction}) and
(\ref{surf}) while using the derivative expansions for the bulk
and brane metrics and curvatures as well as (\ref{Lvsdelta}) and
integrating over the radial direction $\rho$ we get
\begin{equation}
I_{ind}=\int d^d x \sqrt{-\gt}\bigg[\frac{R}{16\pi
G_N}+\frac{\kappa_1}{2\pi}\big(R_{ij}R^{ij}-\frac{d}{4(d-1)}R^2\bigg)+\cdots\bigg]\,.\label{indact2}
\end{equation}
The ellipsis in (\ref{indact2}) are of order
$\mathcal{O}(\partial^6)$ and higher and
\begin{eqnarray}
\frac{1}{G_N}=\frac{2\delta}{d-2}\frac{f'(\mathcal{R}_0)}{G_{d+1}}\,,\qquad
\kappa_1=\frac{\delta^3}{4(d-2)^2(d-4)}\frac{f'(\mathcal{R}_0)}{G_{d+1}}\,,\label{Kappa1}
\end{eqnarray}
with $\mathcal{R}_0$ is given by (\ref{R0}) and we have tuned the
brane tension to be
\begin{equation}
T_{brane}=\frac{d-1}{4\pi\delta G_{d+1}}f'(\mathcal{R}_0)\,.
\end{equation}
Note that all the curvatures in expression (\ref{indact2}) are
constructed from the brane metric $\gt_{ij}$. So far, we have
found the effective Newton constant of the brane $G_N$ in terms of
the bulk gravitational constant $G_{d+1}$. Moreover, we have an
additional parameter $\kappa_1$ on the brane which is expressed in
terms of bulk gravity parameters. It is worth to mention that the
expression (\ref{indact2}) for induced action has the same form as
previously obtained in \cite{Myers:2013lva} for Einstein and
Gauss-Bonnet gravity. However, the effective Newton constant $G_N$
and the coupling $\kappa_1$ have different definitions in terms of
the bulk gravitational couplings.

\section{Entanglement entropy}
\label{EntEnt}

Our goal is to calculate the leading term and the first subleading
term of the entanglement entropy of a general surface with a
smooth geometry in a weakly curved background. Therefore, we
assume a sufficiently large surface with generic geometry on the
RS brane which is weakly curved due to the constraint
(\ref{deltaR}). However, in order to calculate the entanglement
entropy of such a surface, instead of going through QFT
calculations, we are following holographic approach. That is, as
shown in figure (\ref{fig:minsurf}), we extend the general surface
$\widetilde{\Sigma}$ on the brane into the bulk and then for this
bulk surface $\sigma$, we apply holographic entanglement entropy
formula introduced by Ryu and Takayanagi \cite{Ryu:2006bv,
Ryu:2006ef}, i.e.,
\begin{equation}
S_{\sigma}\equiv
\mathrm{min}\frac{\mathcal{A}(\sigma)}{4G_{d+1}}\,.\label{RTformula}
\end{equation}
However, with $f(\mathcal{R})$ gravity, we need to find the
appropriate entropy functional for the bulk surface $\sigma$ and
then extremise the functional to find holographic entanglement
entropy \cite{Hung:2011xb}. A natural guess with a general
covariant Lagrangian $\mathcal{L}(g, R, \nabla R, \cdots)$ would
be the Wald entropy formula \cite{Wald:1993nt, Jacobson:1993vj,
Iyer:1994ys} as following
\begin{equation}
S_{Wald}=-2\pi\int_{\mathrm{\small{hypersurface}}}d^{d-1}X\sqrt{h}\frac{\partial
\mathcal{L}}{\partial
 R_{ijkl}}\hat{\varepsilon}_{ij}\hat{\varepsilon}_{kl}\,,\label{entWald}
\end{equation}
where $\hat{\varepsilon}_{ij}$ is the volume form in the two
dimensional transverse space to the hypersurface. However, this is
known not to be correct in general \cite{Hung:2011xb}. In general,
one must add terms involving the second fundamental forms of the
boundary of $\sigma$. However there is evidence such terms do not
occur for $f(\mathcal{R})$ gravity, e.g., using a novel method
called squashed cone, it has been shown in \cite{Fursaev:2013fta}
that for the bulk action of the $\mathcal{R}^2$ form, which is
specific form of $f(\mathcal{R})$, no extrinsic curvature appears
in the entanglement entropy. Also performing a field redefinition,
one can show that $f(\mathcal{R})$ gravity can be transformed into
a pure Einstein gravity minimally coupled to matter
\cite{Whitt:1984pd}. For the latter, the entropy functional is
simply $A/4G$ and transforming back yields no $\mathcal{K}$ terms.
Therefore, we assume in order to obtain the entropy functional
associated with the bulk surface $\sigma$ for $f(\mathcal{R})$
gravity in the bulk, it is enough to use the Wald entropy formula
(\ref{entWald}). This yields
\begin{equation}
S_{\sigma}=\frac{1}{2G_{d+1}}\int_{\widetilde{\Sigma}}
d^{d-2}y\int_{1}^{\infty} d\rho \sqrt{h}
f'(\mathcal{R})\,,\label{wald}
\end{equation}
where $y^i$ are the coordinates along the entangling surface
$\widetilde{\Sigma}$ and $h_{\alpha\beta}$ is the induced metric
on the codimension-2 surface $\sigma$ with its components are
given as a Taylor series about $\rho=0$ by
\begin{eqnarray}
h_{\rho\rho}=\frac{\delta^2}{4\rho^2}\bigg(1+\ho_{\rho\rho}\rho+\cdots\bigg)\,,\qquad\quad
h_{ab}=\frac{1}{\rho}\bigg(\hz_{ab}+\ho_{ab}\rho+\cdots\bigg)\,.\label{hexpan}
\end{eqnarray}
In the above expression
$\overset{\scriptscriptstyle{(0)}}{h}_{ab}$ is the induced metric
of the entangling surface $\Sigma$ on the AdS boundary and the
coefficients in the expansion are given by \cite{Schwimmer:2008yh}
\begin{equation}
\ho_{\rho\rho}=\frac{\delta^2}{(d-2)^2}K^iK^j\gz_{ij}\,,\qquad\ho_{ab}=\go_{ab}-\frac{\delta^2}{d-2}K^iK^j_{ab}\gz_{ij}\,,\label{hindt}
\end{equation}
where $K^i=\overset{\scriptscriptstyle{(0)}}{h}\,\!^{ab}K^i_{ab}$
is the trace of the second fundamental form of $\Sigma$. It is
worth to clarify that the entangling surface is a codimension-2
with a pair of orthonormal vectors $n^{I}_j$ ($I=0,1$) and
associated extrinsic curvatures $K^{I}_{ab}=\nabla_a n^I_b$.
Contracting with a normal vector gives $K^i_{ab}=n^i_JK^J_{ab}$,
that's why the extrinsic curvatures here and in the following
always carry a coordinate index $i$.

Moreover, since we are using the expansion (\ref{hexpan}) in the
vicinity of the brane at $\rho=1$, then we need to justify that it
converges fast enough to be applicable for finite $\rho$. This
requires not only the background geometry is weakly curved as
indicated in constraint (\ref{deltaR}) but also the entangling
surface $\widetilde\Sigma$ is smooth. Recall that the surface
$\widetilde\Sigma$ is the intersection of the bulk surface
$\sigma$ and the brane at $\rho=1$ with its metric given by
\begin{equation}
\tilde{h}_{ab}=h_{ab}|_{\rho=1}=\overset{\scriptscriptstyle{(0)}}{h}_{ab}+\overset{\scriptscriptstyle{(1)}}{h}_{ab}+\cdots\,.\label{hbrane}
\end{equation}
The derivative expansion (\ref{hbrane}) truncates by imposing that
the characteristic scale of the extrinsic curvatures is small
compared to the AdS scale, i.e.,
\begin{equation}
\delta\,K^i_{ab}\ll1\,,\label{deltaK}
\end{equation}
whereas the same constraint should be held for covariant
derivatives of the extrinsic curvatures which might appear in
higher orders. Again, we are assuming that the expressions are
evaluated in an orthonormal frame.

Now, in order to evaluate the entanglement entropy associated with
$\widetilde\Sigma$, we plug (\ref{hexpan}) into (\ref{wald}),
applying Taylor expansion for $f(\mathcal{R})$ and integrating
over the radial direction $\rho$ from the location of the brane to
infinity we get
\begin{eqnarray}
S_{EE}=\frac{\delta}{2(d-2)G_{d+1}}\int d^{d-2}y \sqrt{\tilde{h}}
f'(\mathcal{R}_0) \bigg[1+\frac{d-2}{2(d-4)}\ho_{\rho\rho}
+\frac{1}{d-4}\hz^{ab}\ho_{ab}+\cdots\bigg]\,,\nonumber\\\label{swald}
\end{eqnarray}
where we have used
\begin{equation}
\sqrt{\hz}=\sqrt{\tilde{h}}\bigg(1-\frac{1}{2}\hz^{ab}\ho_{ab}+\cdots\bigg)\,.\label{h0ht}
\end{equation}
Note that there is no $f''$ term in (\ref{swald}), since curvature
squared term is absent in the derivative expansion
(\ref{rexpansion}).

Finally if we use (\ref{hindt}) along with the expressions in
(\ref{Kappa1}) for the effective Newton constant $G_N$ and
parameter $\kappa_1$ we can rewrite (\ref{swald}) as following
\begin{eqnarray}
S_{EE}=\frac{\mathcal{A}(\widetilde\Sigma)}{4G_N}+\kappa_1\int_{\widetilde\Sigma}
d^{d-2}y
\sqrt{\tilde{h}}\bigg(2R^{ij}\tilde{g}^{\bot}_{ij}-\frac{d}{d-1}R-K^i
K_i\bigg)+\mathcal{O}(\partial^4)\,. \label{sEE}
\end{eqnarray}
It is clear that the leading term is just the area law as it has
been already conjectured in \cite{Bianchi:2012ev}. Again one
should note that the expression (\ref{sEE}) for entanglement
entropy has the same form as previously obtained in
\cite{Myers:2013lva} with Einstein and Gauss-Bonnet gravity in the
bulk. The only distinction is that the effective Newton constant
$G_N$ and the coupling $\kappa_1$ are defined differently in terms
of bulk parameters.

Moreover, the first subleading term can also teach us an
interesting lesson: Let's evaluate the Wald entropy associated to
the entangling surface $\widetilde\Sigma$ by directly applying the
Wald formula (\ref{entWald}) for this surface which is a
codimension-2 hypersurface on the brane with induced action
(\ref{indact2}). Doing so, one obtains
\begin{equation}
S_{\widetilde\Sigma}=\frac{\mathcal{A}(\widetilde\Sigma)}{4G_N}+\kappa_1\int_{\widetilde\Sigma}
d^{d-2}y
\sqrt{\tilde{h}}\bigg(2R^{ij}\tilde{g}^{\bot}_{ij}-\frac{d}{d-1}R\bigg)+\mathcal{O}(\partial^4)\,.
\label{swaldb}
\end{equation}
where we have used the following identities:
\begin{equation}
\hat{\varepsilon}_{ik}\hat{\varepsilon}_j^{~k}=-\tilde{g}_{ij}^{\bot}\,,\qquad\hat{\varepsilon}_{ij}\hat{\varepsilon}^{ij}=-2\,.
\end{equation}
Now comparing (\ref{sEE}) with (\ref{swaldb}), it is evident that
the entanglement entropy for a general surface agrees to the Wald
entropy up to the extrinsic curvature terms. In fact, if the
entangling surface is a {\it Killing horizon}, for which extrinsic
curvatures are vanishing, then both entropies coincide. However,
for a general entangling surface, the Wald entropy does not give
the whole entanglement entropy for the surface; there are some
contributions to the entanglement entropy from non vanishing
extrinsic curvature terms which they do not appear in the Wald
entropy. Indeed, the fact that the entanglement entropy cannot be
completely extracted from the Wald formula has been recently
studied in \cite{Myers:2013lva, Fursaev:2013fta, Dong:2013qoa}.

\section{Discussion}
\label{disc}

In this letter we constructed the RS2 model by taking two copies
of AdS spacetime and gluing them together along a cut-off surface
at some large radius and inserting a brane at this junction.
Therefore, the standard gravity will arise at long distances on
the brane as induced gravity: using a FG expansion around the
brane and integrating out the extra radial direction, we derived
the induced action (\ref{braneaction}) on the brane embedded in
the bulk described by the classical action (\ref{bulkaction})
known as $f(\mathcal{R})$ gravity. As a result, we obtained the
the effective Newton constant $G_N$ and a new coupling $\kappa_1$
on the brane in terms of the bulk parameters in equation
(\ref{Kappa1}). However, since the brane is located at some finite
radial direction, to make sense of the derivative expansion in our
calculations, we demanded background geometry is weakly curved
compared to the AdS scale by imposing constraint (\ref{deltaR}).

To calculate the entanglement entropy associated with a general
surface $\widetilde\Sigma$ on the brane, we used the RT
holographic prescription (\ref{RTformula}) in the context of the
RS2 model. In particular, we argued that the entropy functional
(\ref{wald}) associated to the bulk surface $\sigma$, is the one
we need to extremise in the RT formula (\ref{RTformula}) for
$f(\mathcal{R})$ gravity in the bulk. Note that the holographic
surface $\sigma$ is an extension of the surface $\widetilde\Sigma$
into the bulk. Therefore, in order to obtain the EE of
$\widetilde\Sigma$, we need to integrate over the extra
holographic direction. Again, we used derivative expansion to
integrate out the radial coordinate and to ensure the convergence
of the expansion, we demanded not only a smooth geometry for the
ambient metric but also for the entangling surface. In other word,
along with a constraint (\ref{deltaR}) for the intrinsic curvature
of the boundary metric, we imposed constraint (\ref{deltaK}) for
the extrinsic curvature of the entangling surface on the boundary.

Carrying all the calculations, we finally obtained expression
(\ref{sEE}) for a general entangling surface indicating that EE of
any region surrounded by a smooth entangling surface is {\it
finite} and the leading contribution is given precisely by the
Bekenstein-Hawking area law \cite{Bekenstein:1972tm,
Bekenstein:1973ur, Bekenstein:1974ax, Hawking:1971vc}. Hence this
model confirmed a conjecture by Bianchi and Myers
\cite{Bianchi:2012ev}. We also calculated the first leading
corrections to the area law and found that EE coincides with the
Wald entropy if the entangling surface is a Killing horizon but
for a general surface in addition to the Wald entropy, there are
terms dependent on the extrinsic curvature of the entangling
surface. It is easy to see that in the spacial case where
$f(\mathcal{R})=\mathcal{R}$, which corresponds to have an
Einstein gravity in the bulk, we reproduce all the results
previously obtained in \cite{Myers:2013lva}. Furthermore, we get
the same form of $R^2$ term in the action and $R-K^2$ term in the
EE as we previously obtained in \cite{Myers:2013lva}, however,
with different gravity theory in the bulk. In fact, the type of
theory in the bulk manifests itself just in the effective Newton
constant $G_N$ and the coupling $\kappa_1$ which are expressed in
terms of bulk parameters.

\section*{Acknowledgement}

Many thanks to Rob Myers, Misha Smolkin, Nima Doroud and Sergey
Solodukhin for encouragement and useful discussions. Research at
Perimeter Institute is supported by the Government of Canada
through Industry Canada and by the Province of Ontario through the
Ministry of Research \& Innovation. I also acknowledge support
from Ontario Graduate Scholarship and President's Graduate
Scholarship awarded by University of Waterloo 2013-2014.

\bibliography{fR}{}
\bibliographystyle{JHEP}

\end{document}